\documentclass[epj]{svjour}
\usepackage{graphics}

\newcommand{\Pe}{\mathrm e}
\newcommand{\Pep}{\mathrm {e^+}}
\newcommand{\Pem}{\mathrm {e^-}}
\newcommand{\Pne}{\nu_{\mathrm{e}}}
\newcommand{\Pt}{\mathrm t}

\newcommand{\PW}{\mathrm W}
\newcommand{\PZ}{\mathrm Z}
\newcommand{\PH}{\mathrm H}
\newcommand{\MW}{M_\PW}
\newcommand{\MZ}{M_\PZ}
\newcommand{\MH}{M_\PH}
\newcommand{\Mt}{m_\Pt}
\newcommand{\Oa}{{{\cal{O}}(\alpha)}}

\newcommand{\eennh}{\Pep\Pem\to\nu\bar\nu\PH}
\newcommand{\eetth}{\Pep\Pem\to\Pt\bar\Pt\PH}
\newcommand{\GF}{ {G_\mu}}
\newcommand{\GeV}{\unskip\,\mathrm{GeV}}
\newcommand{\TeV}{\unskip\,\mathrm{TeV}}
\newcommand{\fb}{\unskip\,\mathrm{fb}}
\newcommand{\gsim}
{\mathrel{\raisebox{-.3em}{$\stackrel{\displaystyle >}{\sim}$}}}

\def\reffi#1{Figure~\ref{#1}}
\def\citere#1{Ref.~\cite{#1}}
\def\citeres#1{Refs.~\cite{#1}}

\begin{document}

\title{Electroweak Corrections to  
$\mathbf{e^+ e^- \to \nu \bar{\nu} H}$ and 
$\mathbf{e^+ e^- \to t \bar{t} H}$%
\thanks{\emph{This work was supported in part by
the Swiss Bundesamt f\"ur Bildung und Wissenschaft and by the European
Union under contract HPRN-CT-2000-00149.}}%
}
\author{A.\ Denner\inst{1}, S.\ Dittmaier\inst{2}, 
\underline{M.\ Roth}\inst{2} \and M.\ M.\ Weber\inst{1}%
}                  
\institute{
Paul Scherrer Institut, W\"urenlingen und Villigen,
CH-5232 Villigen PSI, Switzerland
\and
Max-Planck-Institut f\"ur Physik 
(Werner-Heisenberg-Institut),
D-80805 M\"unchen, Germany 
}
\date{Received: date / Revised version: date}
\abstract{
The most interesting Higgs-production processes at future 
$\Pep \Pem$ colliders belong to the process class 
$\Pep \Pem \to f \bar{f} \PH$. 
We study the full $\Oa$ corrections to this reaction in the Standard 
Model for neutrinos and top quarks in the final state. 
Leading higher-order corrections from initial-state radiation 
and QCD corrections are also taken into account.
Although cancellations between the different kinds of 
corrections occur, the full corrections are of the order 
of $\pm 10\%$ and thus important ingredients in the theoretical 
predictions for future $\Pep \Pem$ colliders.
}

\maketitle

\markboth
{A.\ Denner, S.\ Dittmaier, M.\ Roth, M.\ M.\ Weber: 
Electroweak Corrections to $\eennh$ and $\eetth$}
{A.\ Denner, S.\ Dittmaier, M.\ Roth, M.\ M.\ Weber: 
Electroweak Corrections to $\eennh$ and $\eetth$}

\section{Introduction}
\label{intro}

One of the most important future challenges in particle physics is the 
understanding of the mechanism of electroweak symmetry breaking 
and the discovery of the up to now only missing particle of the Electroweak 
Standard Model (SM), the Higgs boson. The mass of the Higgs boson 
is expected to be in the range between the current lower experimental 
bound of $114.4\GeV$ and $1\TeV$, where a light Higgs boson (with 
$\MH\sim100$--$200\GeV$) is favoured by the global fit of the SM 
to electroweak precision data.
While the LHC will find a SM Higgs boson in the full mass 
range up to $1\TeV$ if it exists and has no exotic properties, a complete 
determination of the Higgs interactions is only possible in the clean 
environment of an $\Pep \Pem$ collider. 

In this work we focus on the reactions
$\Pep \Pem \to f \bar{f} \PH$ where $f$ is a
neutrino or a top quark, which belong to the most 
interesting Higgs production processes at future 
$\Pep \Pem$ colliders. 

\section{Radiative corrections to 
$\mathbf{e^+ e^- \to \nu \bar{\nu} H}$}
\label{sec:eennh}

\begin{figure*}
\centerline{
\resizebox{0.44\textwidth}{!}{\includegraphics{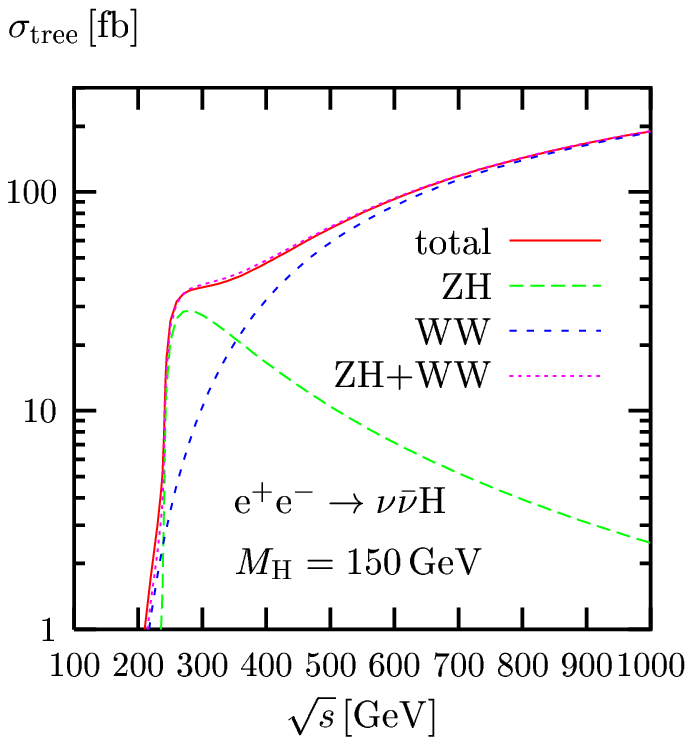}}
\resizebox{0.44\textwidth}{!}{\includegraphics{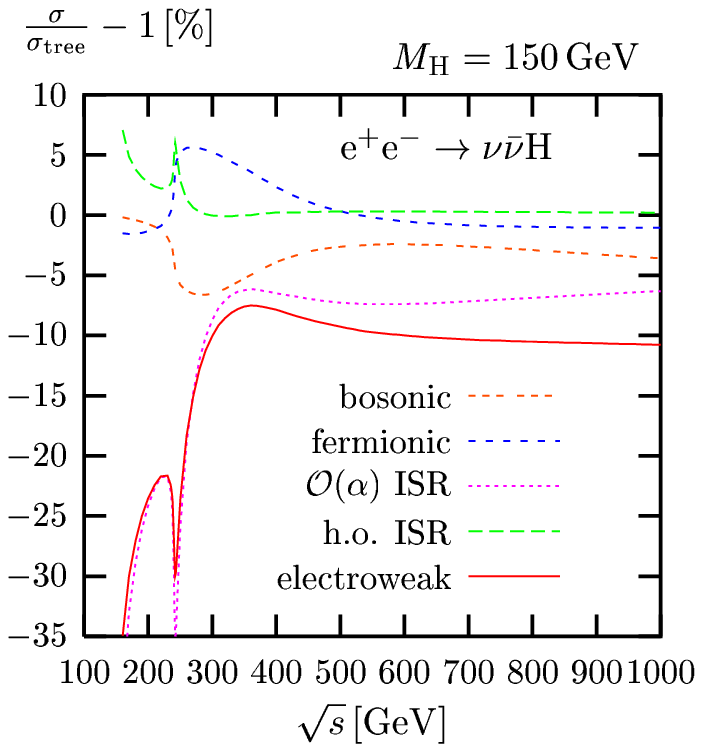}}
}
\caption{Lowest-order cross section and contributions from ZH-production 
channel, WW-fusion channel, and their sum (l.h.s.) as well as 
relative corrections (r.h.s.) in the $\GF$ scheme for a Higgs-boson mass 
$\MH=150\GeV$}
\label{fig:eennh}     
\end{figure*}

At an $\Pep \Pem$ collider the two main Higgs production 
mechanisms in the SM are Higgs radiation off Z bosons, so-called 
Higgs strahlung, and Higgs production via WW fusion.
Both mechanisms are present in the reaction 
$\Pep \Pem \to \nu_l \bar{\nu}_l \PH$ where
$l$ can be an $\Pe$, $\mu$, or $\tau$. The l.h.s.\ of  
\reffi{fig:eennh} shows the different contributions to the lowest-order
cross section as a function of the centre-of-mass (CM) energy $\sqrt{s}$
for $\MH=150\GeV$ (see \citere{Denner:2003yg} for details).
While the Higgs-strahlung contribution to the total cross section 
rises sharply at threshold to a maximum  of a few tens of GeV above 
$\sqrt{s}=\MZ+\MH$ and falls off as $1/s$,
the WW-fusion channel, which is only present in the reaction 
$\Pep \Pem \to \Pne \bar\Pne \PH$,
dominates the cross section well above the ZH threshold and grows 
as $\ln s$ in the high-energy limit. The difference between 
the total cross section and the sum ZH+WW is the interference 
between both channels which is relatively small.

The $\Oa$ electroweak corrections to the process $\Pep \Pem \to \PZ \PH$
have been calculated in 
\citeres{Fleischer:1982af,Kniehl:1991hk,Denner:1992bc}, 
and a Monte Carlo algorithm for the calculation of the real photonic 
corrections to this process was described in \citere{Berends:dw}. 
The electroweak corrections to the full process 
$\Pep \Pem \to \nu \bar{\nu} \PH$ have attracted a lot of interest recently. 
Analytical results for the one-loop corrections to this process have been 
studied in \citere{Jegerlehner:2002es}; however, no numerical results are 
given there. The contributions of fermion and sfermion loops in the 
Minimal Supersymmetric Standard Model (MSSM) have been evaluated in
\citeres{Eberl:2002xd,Hahn:2002gm}.
Complete calculation of the full $\Oa$ 
electroweak corrections to $\Pep \Pem \to \nu \bar{\nu} \PH$
in the SM have been performed in \citeres{Denner:2003yg,Belanger:2002ik}.  
These calculations agree within $0.3\%$, which is of the same order as the 
integration error of \citere{Belanger:2002ik}.

In the following we briefly summarize some results of our 
calculation \cite{Denner:2003yg} of the $\Oa$ electroweak corrections,
where we included also corrections from initial-state 
radiation (ISR) beyond $\Oa$ in the structure-function approach. 
The calculation is done in the $\GF$ scheme which absorbs the 
corrections proportional to $\Mt^2/\MW^2$ in the fermion--W-boson couplings 
and the running of $\alpha(Q^2)$ from $Q^2=0$ to the electroweak scale. 
The numerical evaluation of the virtual corrections is particularly 
complicated due to the appearance of pentagon diagrams. Therefore, we 
have applied the approach of \citere{Denner:2002ii} which avoids the 
appearance of inverse Gram determinants in the reduction of 
tensor 5-point functions. 
The soft and collinear singularities are treated both in the dipole 
subtraction method following \citeres{Dittmaier:1999mb,Roth:1999kk} 
and in the phase-space slicing method following closely \citere{Bohm:1993qx}. 
Two completely independent Monte Carlo programs have been constructed; 
one applies the multi-channel approach similar to 
\citeres{Roth:1999kk,Denner:1999gp,Dittmaier:2002ap}, the second uses 
{\sl VEGAS} \cite{Lepage:1977sw}. 

The relative corrections are shown on the r.h.s.\ of \reffi{fig:eennh}. The
ISR corrections vary strongly in the region of the $\PZ\PH$ threshold
but are nearly flat for energies above $400\GeV$.
They are always negative since the lowest-order cross section is
continuously rising. The fermionic corrections reach a maximum of
about $6\%$ in the region where the $\PZ\PH$-production channel 
dominates and are small above $500\GeV$ where the WW-fusion process 
is most important. The non-ISR bosonic corrections exhibit a minimum of about 
$-7\%$ at \mbox{$\sim300\GeV$} and are between $0\%$ and $-4\%$ elsewhere. 
Near the $\PZ\PH$ threshold the electroweak corrections are dominated 
by the $\Oa$ ISR corrections from ZH production, while for 
higher energies, where the main contributions come from WW fusion, 
the corrections become flat reaching relative 
corrections of about $-10\%$ above $500\GeV$.

\section{Radiative corrections to 
$\mathbf{e^+ e^- \to t \bar{t} H}$}
\label{sec:eetth}

\begin{figure*}
\centerline{
\resizebox{0.44\textwidth}{!}{\includegraphics{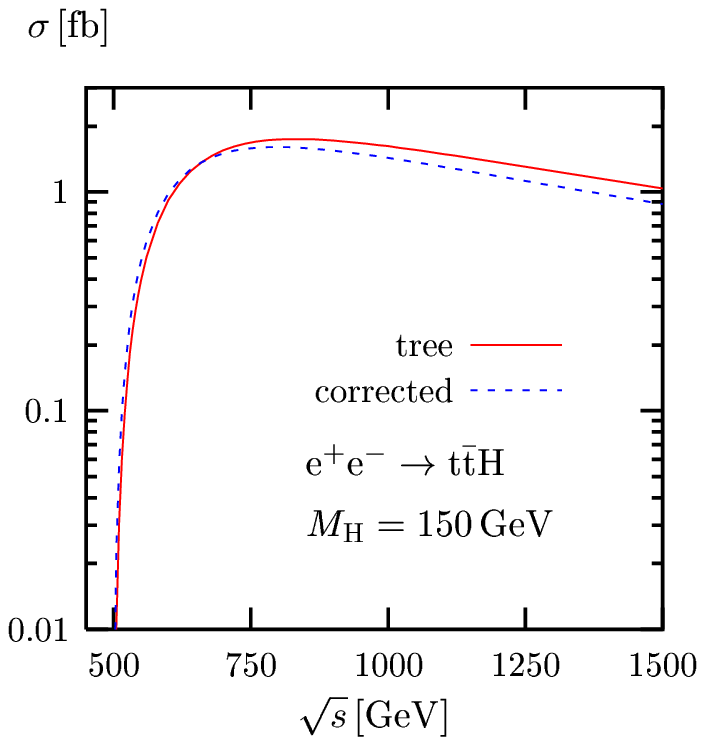}}
\resizebox{0.44\textwidth}{!}{\includegraphics{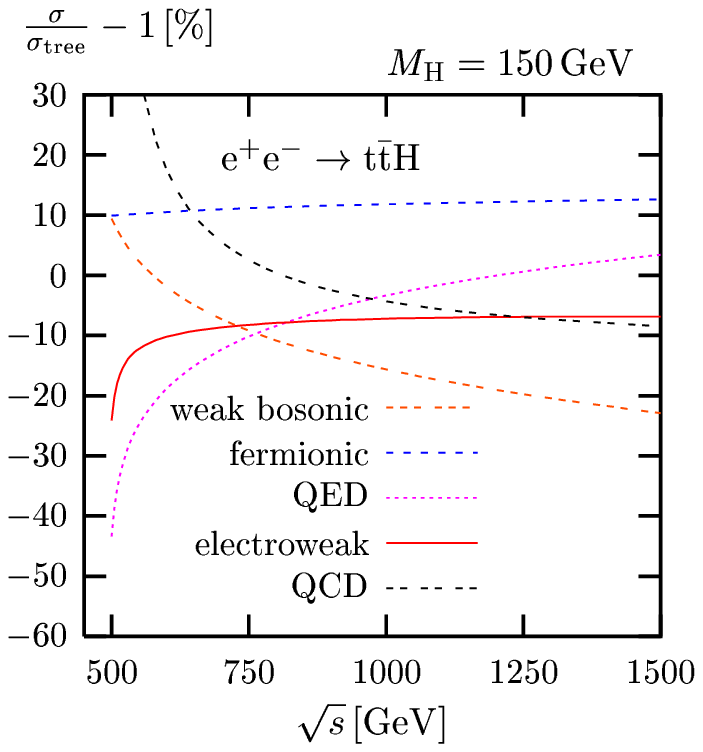}}
}
\caption{Lowest-order and corrected cross section (l.h.s.) as well as
relative corrections (r.h.s.) in the $\GF$ scheme for a Higgs-boson mass 
$\MH=150\GeV$}
\label{fig:eetth}     
\end{figure*}

Another interesting Higgs production process is the reaction 
$\Pep \Pem \to \Pt \bar\Pt \PH$.
If the Higgs-boson mass is not too large, i.e.\ 
$\MH\sim100$--$200\GeV$, the Higgs boson is produced
mainly through Higgs radiation off top quarks, while emission from 
intermediate Z bosons plays only a minor role.
Therefore, this process can be used for the determination of the   
top-quark Yukawa coupling, $g_{\Pt\bar\Pt\PH}$, which 
is by far the largest one in the SM ($g_{\Pt\bar\Pt\PH}\approx 0.5$).
If the Higgs boson is light, i.e.\ $\MH\sim120\GeV$, a precision of 
around $5\%$ can be reached at an $\Pep \Pem$ collider operating at 
$\sqrt{s}=800\GeV$ with a luminosity of 
$\int L\,\mathrm{d} t \sim1000\fb^{-1}$ \cite{Baer:1999ge}.  
Combining the $\Pt \bar\Pt \PH$ channel with 
information from other Higgs-production and decay processes an even 
better accuracy can be obtained in a combined fit \cite{Battaglia:2000jb}.  

The ${\cal O}(\alpha_s)$ corrections to the total cross section within 
the SM have been calculated for the dominant photon-exchange channel 
in \citere{Dawson:1998ej}, while the full set of diagrams has been 
evaluated in \citere{Dittmaier:1998dz}. 
Recently, considerable progress has been achieved in the calculation of 
the electroweak corrections to $\eetth$. Results for the electroweak 
$\Oa$ corrections in the SM have been presented in
\citeres{You:2003zq,Belanger:2003nm,Denner:2003ri}. 
While the results of \citeres{Belanger:2003nm,Denner:2003ri} agree well, 
those of \citere{You:2003zq} differ at large CM energies and close to 
threshold. 

In the last part of this article we show some results of our 
calculation of the $\Oa$ electroweak corrections and
of the ${\cal O}(\alpha_s)$ corrections. Although the virtual corrections 
are much more complex than for $\eennh$, 
we were able to perform the calculation using the same computational
techniques as in the former case. Results for total cross sections and 
various distributions have been presented in \citere{Denner:2003ri}. 

On the l.h.s.\ of \reffi{fig:eetth} we show the lowest-order cross section 
and the cross section including both electroweak and QCD 
corrections as a function of the CM energy for $\MH=150\GeV$. 
Away from the kinematic threshold at $\sqrt{s}=2 \Mt+\MH$
the total cross section is typically of the order of a few $\fb$ and 
becomes maximal at an energy of about $800\GeV$. 
The relative corrections are presented on the r.h.s.\ of \reffi{fig:eetth}. 
While the weak bosonic corrections are around 
$+10\%$ close to threshold and fall off rapidly with increasing CM 
energy, the fermionic corrections are about $+10\%$ and depend only 
weakly on the CM energy. For energies above $600\GeV$, the fermionic and 
weak bosonic contributions partially cancel. 
The QED corrections, which include both the complete photonic and 
higher-order ISR corrections, are about $-40\%$ at threshold and rise 
to a few per cent at $1.5\TeV$. In the electroweak corrections, 
both QED and weak contributions partially compensate each other. 
The QCD corrections are positive and rather large in the threshold region, 
where soft-gluon exchange between in the $\Pt\bar\Pt$ system leads to a 
Coulomb-like singularity. In the region above threshold, the QCD corrections 
decrease and even turn negative for energies $\gsim800\GeV$. 

In summary, for both processes $\eennh$ and $\eetth$ 
we find corrections that are typically of the order of 
$\pm 10\%$ and thus important ingredients in the theoretical 
predictions for future $\Pep \Pem$ colliders.

\end{document}